\documentclass[conference]{IEEEtran}

\usepackage{graphicx}
\usepackage{epstopdf}
\usepackage{standalone}
\usepackage[nolist ]{acronym}
\usepackage{tcolorbox}

\usepackage{pdfpages}
\usepackage{tikz}
\usetikzlibrary{positioning}
\usepackage[draft]{hyperref}
\usepackage{pdfcomment}
\usepackage{hologo}
\usepackage{flushend}

\usepackage{xstring}

\usepackage[utf8]{inputenc}
\usepackage{marvosym}

\usepackage{algorithm}
\usepackage{algpseudocode}
\usepackage{tikz}

\usepackage{listings}
\definecolor{ListingBackground}{rgb}{0.97,0.97,0.97}
\usepackage{pgfplots}
\pgfplotsset{compat=newest}
\pgfplotsset{
    box plot/.style={
        /pgfplots/.cd,
        fill=blue!30,
        only marks,
        mark=-,
        mark size=0.2em,
        /pgfplots/error bars/.cd,
        y dir=plus,
        y explicit,
    },
    box plot box/.style={
        /pgfplots/error bars/draw error bar/.code 2 args={%
            \draw  ##1 -- ++(.2em,0pt) |- ##2 -- ++(-.2em,0pt) |- ##1 -- cycle;
        },
        /pgfplots/table/.cd,
        y index=2,
        y error expr={\thisrowno{3}-\thisrowno{2}},
        /pgfplots/box plot
    },
    box plot top whisker/.style={
        /pgfplots/error bars/draw error bar/.code 2 args={%
            \pgfkeysgetvalue{/pgfplots/error bars/error mark}%
            {\pgfplotserrorbarsmark}%
            \pgfkeysgetvalue{/pgfplots/error bars/error mark options}%
            {\pgfplotserrorbarsmarkopts}%
            \path ##1 -- ##2;
        },
        /pgfplots/table/.cd,
        y index=4,
        y error expr={\thisrowno{2}-\thisrowno{4}},
        /pgfplots/box plot
    },
    box plot bottom whisker/.style={
        /pgfplots/error bars/draw error bar/.code 2 args={%
            \pgfkeysgetvalue{/pgfplots/error bars/error mark}%
            {\pgfplotserrorbarsmark}%
            \pgfkeysgetvalue{/pgfplots/error bars/error mark options}%
            {\pgfplotserrorbarsmarkopts}%
            \path ##1 -- ##2;
        },
        /pgfplots/table/.cd,
        y index=5,
        y error expr={\thisrowno{3}-\thisrowno{5}},
        /pgfplots/box plot
    },
    box plot median/.style={
        /pgfplots/box plot
    },
    boxplot/every median/.style={
    	ultra thick,dashed,cyan
    }
}

\definecolor{flexicolor}{RGB}{46,49,146}
\definecolor{amaricolor}{RGB}{237,28,36}

\usepackage{xspace}

\usepackage[binary-units=true]{siunitx}
\usepackage{psfrag}
\usepackage{graphicx}
\usepackage{tabularx,booktabs}
\usepackage{multirow}
\usepackage{rotating}

\renewcommand{\baselinestretch}{1}

\usepackage{multirow}

\usepackage[cmex10]{amsmath}
\usepackage[caption=false,font=footnotesize]{subfig}
\usepackage{stfloats}
\begin{document}

\newcommand{\paperTitle}{Unmanned Aerial Vehicles in Logistics: Efficiency Gains and Communication Performance of Hybrid Combinations of Ground and Aerial Vehicles}
\newcommand{\paperAuthors}{Manuel Patchou, Benjamin Sliwa, and Christian Wietfeld}
\newcommand{\paperEmails}{$\{$Manuel.Mbankeu, Benjamin.Sliwa, Christian.Wietfeld$\}$@tu-dortmund.de}

\newcommand{\figurePadding}{0pt}
\newcommand{\figureTopPadding}{\figurePadding}
\newcommand{\figureBottomPadding}{\figurePadding}

\newcommand{\dummy}[3]
{
	\begin{figure}[b!]  
		\begin{tikzpicture}
		\node[draw,minimum height=6cm,minimum width=\columnwidth]{\LARGE #1};
		\end{tikzpicture}
		\caption{#2}
		\label{#3}
	\end{figure}
}

\newcommand{\dummyhere}[3]
{
	\begin{figure}[h!]  
		\begin{tikzpicture}
		\node[draw,minimum height=6cm,minimum width=\columnwidth]{\LARGE #1};
		\end{tikzpicture}
		\caption{#2}
		\label{#3}
	\end{figure}
}

\newcommand{\wDummy}[3]
{
	\begin{figure*}[b!]  
		\begin{tikzpicture}
		\node[draw,minimum height=6cm,minimum width=\textwidth]{\LARGE #1};
		\end{tikzpicture}
		\caption{#2}
		\label{#3}
	\end{figure*}
}

\newcommand{\basicFig}[7]
{
	\begin{figure}[#1]  	
		\vspace{#6}
		\centering		  
		\includegraphics[width=#7\columnwidth]{#2}
		\caption{#3}
		\label{#4}
		\vspace{#5}	
	\end{figure}
}
\newcommand{\fig}[4]{\basicFig{#1}{#2}{#3}{#4}{0cm}{0cm}{1}}

\newcommand{\subfig}[3]
{
	\subfloat[#3]
	{
		\includegraphics[width=#2\textwidth]{#1}
	}
	\hfill
}

\newcommand{\subtikz}[3]
{
	\subfloat[#3]
	{
		\resizebox{#2}{!}{\input{#1}}
	}
}

\newcommand\circled[1] % caution with using in captions: \protect \circled
{
	\tikz[baseline=(char.base)]
	{
		\node[shape=circle,draw,inner sep=1pt] (char) {#1};
	}\xspace
}

\newcommand\truckonly{\textbf{truck only}\xspace}
\newcommand\onsite{\textbf{on site}\xspace}
\newcommand\enroute{\textbf{en route}\xspace}
\begin{acronym}

	\acro{LTE}{Long Term Evolution}	
	\acro{C-V2X}{Cellular Vehicle-to-Everything}
	\acro{UE}{User Equipment}
	\acro{eNB}{evolved Node B}
	\acro{RSRP}{Reference Signal Received Power}
	\acro{CNI}{Communication Networks Institute}
	\acro{GCS}{Ground Control Station}
	\acro{CAM}{Cooperative Awareness Message}
	\acro{IPC}{Interprocess Communication}
	
	\acro{UAV}{Unmanned Aerial Vehicle}
	\acro{UAS}{Unmanned Aerial System}
	\acro{ITS}{Intelligent Transportation System}
	\acro{ns-3}{Network Simulator 3}
	\acro{LIMoSim}{Lightweight ICT-centric Mobility Simulation}
	\acro{OSM}{OpenStreetMap}
	
	\acro{MIP}{Mixed Integer Programming}
	\acro{PDR}{Packet Delivery Ratio}
	
	\acro{FSTSP}{Flying Sidekick Travelling Salesman Problem}
	\acro{TSP}{Travelling Salesman Problem}
	\acro{TSP-D}{Travelling Salesman Problem with Drone}
	\acro{VRP}{Vehicle Routing Problem}
	\acro{PDP}{Pickup and Delivery}
	
\end{acronym}

\newcommand\uav{\ac{UAV}\xspace}
\newcommand\uavs{\acp{UAV}\xspace}
\newcommand\uas{\ac{UAS}\xspace}
\newcommand\uass{\acp{UAS}\xspace}
\newcommand\its{\ac{ITS}\xspace}
\newcommand\itss{\acp{ITS}\xspace}

\acresetall
\title{\paperTitle}

\author{\IEEEauthorblockN{\textbf{\paperAuthors}}
	\IEEEauthorblockA{Communication Networks Institute,	TU Dortmund University, 44227 Dortmund, Germany\\
		e-mail: \paperEmails}}

\maketitle

%
% Make your adjustments here
%
\def\COPYRIGHTYEAR{2019}
\def\CONFERENCE{IEEE Vehicular Networking Conference (VNC) 2019} % set after the paper has been accepted

\def\bibtex
{
@InProceedings\{Patchou/etal/2019a,
	author    = \{Manuel Patchou and Benjamin Sliwa and Christian Wietfeld\},
	title     = \{Unmanned aerial vehicles in logistics: \{E\}fficiency gains and communication performance of hybrid combinations of ground and aerial vehicles\},
	booktitle = \{IEEE Vehicular Networking Conference (VNC)\},
	year      = \{2019\},
	address   = \{Los Angeles, USA\},
	month     = \{Dec\},
\}
}
\ifx\CONFERENCE\VOID
\def\conferencenotice{Submitted for publication}
\def\copyrightnotice{}
\else
\ifx\DOI\VOID
\def\conferencenotice{Accepted for presentation in: \CONFERENCE}	
\else
\def\conferencenotice{Published in: \CONFERENCE\\DOI: \href{http://dx.doi.org/\DOI}{\DOI}

}
\fi
\def\copyrightnotice{
	\copyright~\COPYRIGHTYEAR~IEEE. Personal use of this material is permitted. Permission from IEEE must be obtained for all other uses, including reprinting/republishing this material for advertising or promotional purposes, collecting new collected works for resale or redistribution to servers or lists, or reuse of any copyrighted component of this work in other works.
}
\fi
\def\overlayimage{%
	\begin{tikzpicture}[remember picture, overlay]
	\node[below=5mm of current page.north, text width=20cm,font=\sffamily\footnotesize,align=center] {\conferencenotice \vspace{0.3cm} \\ \pdfcomment[color=yellow,icon=Note]{\bibtex}};
	\node[above=5mm of current page.south, text width=15cm,font=\sffamily\footnotesize] {\copyrightnotice};
	\end{tikzpicture}%
}
\overlayimage
\begin{abstract}
	
%
% Motivation
%
\acp{UAV} have drastically gained popularity in various \ac{ITS} applications to improve the safety and efficiency of transportation systems. In this context, the combination of ground vehicles, such as delivery trucks, with drones to assist in the last-mile pick-up and delivery of the parcels has been recently proposed.
%
% Challenge
%
While aerial vehicles promise increased efficiency based on flexible routes and parallelized operation, highly reliable wireless communication is also required for the control and coordination of potentially many drones acting in a self-organized way.
%
% Approach
%
In this paper, we analyze the improvements procured by drone usage in parcel delivery compared to traditional delivery and propose a simulation framework to further quantify the efficiency gains of the parcel delivery logistics and to analyze the performance of different wireless communications options.
To this end, we consider a heterogeneous vehicle routing problem with various constraints. We consider two approaches regarding the dispatching and recovery of drones and evaluate their benefits as opposed to parcel delivery with a classic truck only.
%
% Network
%
Furthermore, we compare two networking technologies for enabling coordination of the self-organizing teams of drones with a realistically modeled environment: one approach relying on base station oriented \ac{LTE} vs. a more decentralized \ac{C-V2X} solution.
%
% Results
%
The results show time savings of nearly 80\% can be achieved through drone usage and that the negative impact of urban shadowing on network communications in the base station oriented \ac{LTE} approach can be compensated by leveraging decentralized \ac{C-V2X} communications.

\end{abstract}

% Simulative evaluation
% C-V2X, IEEE 802.11p

\IEEEpeerreviewmaketitle

\section{Introduction}

%
% Introduction: UAVs in near-field delivery
% 
As transport is one of the main pillars of modern economy, last mile delivery has become fundamental to the E-Commerce industry \cite{Aurambout/etal/2019a} thus
requiring new solutions in order to face the challenges of \ac{PDP}.
Integrating \acp{UAV} into \acp{ITS} yields tremendous potential regarding the development of new applications and services \cite{Menouar/etal/2017a} as they offer more flexibility in motion than conventional terrestrial vehicles.
Most of these novel applications are enhancements of already existing ones by deploying autonomous vehicles to decrease operation costs.
Such is the case of \uav-empowered \ac{PDP} which consists of carrying out parcel delivery logistics with a truck aided by a fleet of drones .

%
% Problem statement
%
Parcel delivery with one vehicle already yields a non negligible complexity.
Therefore, enhancing it with autonomous \uav support adds to the preexisting logistics complexity \cite{Sliwa/etal/2019b} and threatens to tone down the benefits of \uav assistance in this application context.
Several studies were conducted to ascertain the logistics benefits of \uav assistance in parcel delivery scenarios and proof-of-concept implementations have already been introduced \cite{Carlsson/Song/2018a}.
They formulate the drone-aided parcel distribution as an optimization problem to which they suggest heuristic approaches \cite{Marinelli2018}.\\
Furthermore, being an \its application, the performance of \uav-enhanced near-field delivery heavily depends on a suitable communication infrastructure.
There are still few studies investigating this specific \its use case with its dependency to network communications.\\
The integration of \acp{UAV} into existing and upcoming cellular networks is an active field of research \cite{Lin/etal/2018a} as it is expected to facilitate wireless broadcast and support high rate transmissions \cite{Li/etal/2019a}. It also requires the use of anticipatory networking in order to forecast the evolution of network conditions and offer the best achievable connectivity at any times \cite{Bui/etal/2017}.\\
A comprehensive summary of the recent
advances in \uav communications and the challenges they bring about,
with an emphasis on how to integrate \uavs into the forthcoming
fifth-generation (5G) and future cellular networks is provided by \cite{Zeng/etal/2019a}.

%
% Solution approach vs state of the art
%
In this paper we investigate the benefits of \uav-empowered near-field delivery by analyzing the influence of various \uav deployment schemes on the total completion time of the delivery tour and the mean energy consumption per drone. We further compare the
performance of two network communication technologies in the parcel delivery application context.
%
% Structure of the paper
%
This paper provides the following contributions:
\begin{itemize}
	\item Logistics \textbf{simulation} of \textbf{\uav-empowered} \ac{PDP}.
	\item Comparison of \textbf{communication technologies} \ac{LTE} and \ac{C-V2X} in parcel delivery application context.
\end{itemize}
%
% Structure of the paper
%
The paper is structured as follows: After discussing state of the art approaches in Sec.~\ref{sec:related_work}, we present the proposed \ac{UAV}-empowered \ac{PDP} and introduce the methodological setup for the evaluation campaign in Sec.~\ref{sec:approach}. Finally, the results of different evaluations are presented and discussed in Sec.~\ref{sec:results}.

\section{Related Work}
\label{sec:related_work}

%
% UAV Delivery problems
%
The idea of enhancing parcel delivery with drones is currently trending and several implementation approaches have already been discussed. These all entail solving the problem of finding the optimal route for a delivery truck, taking into account the mobility of its associated \uavs, thus enabling them to autonomously deliver parcels up to a certain distance from the truck.

%
% FSTSP
%
One of the problem's earliest introductions \cite{Murray2015} labeled it as the \ac{FSTSP} and introduced a \ac{MIP} formulation and a heuristic approach. Their heuristic begins with an initial \ac{TSP} tour and iteratively considers whether a node should be served by a drone or not.
%
% Other FSTSP and TSP
%
The \ac{FSTSP} is handled with different approaches in \cite{Agatz2018} \cite{Wang2017} to minimize the total traveling time.
%
% Energy optimizing
%
A cost function that considers an energy consumption model and drone re-use is therefore proposed in 
\cite{Dorling2017}. The importance of reusing
drones and optimizing battery size in drone delivery routing problems is confirmed.
%
% En route drone dispatching and recovery
%
As far as our observation goes, all of the mentioned studies make a basic assumption: drone dispatching and recovery must take place at a delivery node.
\cite{Marinelli2018}
presents a novel approach aiming to maximize the drone usage in parcel delivering. The authors consider that a truck can deliver and pick up a drone not only at a delivery target but also along a route arc.

%
% UAVs in ITS
%
The key role played by \uavs in enabling the next generation of \itss \cite{Menouar/etal/2017a} has induced research momentum regarding \uav related \its applications.
%
% LIMoSim
%
For modeling such applications, 
the lightweight vehicular simulation framework \ac{LIMoSim} \cite{Sliwa/etal/2017a} was developed following a shared codebase approach to allow a seamless integration with other simulators, in contrast to the state-of-the art \ac{IPC}-based coupling methods.
It focuses on highly dynamic traffic scenarios where decision processes and routes are determined at runtime \cite{Sliwa/etal/2017a}.
\ac{LIMoSim} is further extended in \cite{Sliwa_etal_2019c} with \uav mobility and an integration with \ac{ns-3} to support simulations of aerial and ground-based vehicular networks.
As the mobility is simulated in a discrete event fashion, \ac{LIMoSim} integrates particularly well with discrete event network simulations.
The shared code base approach allows to natively bring together mobility and network simulators. This enables the development of novel context-aware optimization methods that exploit interdependencies between both domains.

%\ac{LIMoSim} \cite{Sliwa/etal/2017a} which is a lightweight and extensible framework providing a native vehicular simulation and shared code integration to external network simulators such as \ac{ns-3} \cite{Henderson/etal/2008a} was developed.
%Mobility and communication are natively brought together to enable the development of novel context-aware optimization methods that exploit interdependencies between both domains.
%\ac{LIMoSim} is further extended in \cite{Sliwa_etal_2019c} with \uav mobility and an integration with \ac{ns-3} to support simulations of aerial and ground-based vehicular networks.
%
%
%
A realistic model for simulating communications between \uavs and ground vehicles is proposed in \cite{Hadiwardoyo/etal/2019b}. It supports the usage of mobile infrastructure to broadcast alerts in situations of emergency.
The use of \uavs to build an aerial-ground cooperative vehicular networking architecture is proposed in \cite{Zhou/etal/2015a}.
Multiple \uavs, forming an aerial sub-network, aid the ground vehicular sub-network through air-to-air (A2A) and air-to-ground (A2G) communications. \uavs can be dispatched to areas of interest to collect data and share it with ground vehicles.
The flexible mobility of \uavs while being one of their key features is also a challenging point for their integration in the communication infrastructure.
%
% B.A.T. Mobile
%
In \cite{Sliwa/etal/2016b}, 
an approach which leverages application layer knowledge derived from mobility control algorithms guiding the behavior of \uavs to fulfill a dedicated task is presented. It thereby enables the integration of a \uav trajectory prediction with the routing protocol to avoid unexpected route breaks and packet loss.
Finally, the problem of \uav docking station placement for ITS is investigated in \cite{Ghazzai/etal/2017a}. The objective is to determine the best locations for a given number of docking stations that the operator aims to install in a large geographical area.

\section{Simulation of UAV-enabled Pickup and Delivery}
\label{sec:approach}

In this section, we give a detailed presentation of the methods and tools which were used to carry out our analysis.

\subsection{Simulation Environment}
%
% Joint simulation LIMoSim and Network simulation with ns3
%

To model \uav-assistance in near-field delivery, a suitable simulation environment supporting hybrid vehicle scenarios is required.
The platform devised in \cite{Sliwa_etal_2019c},
which relies on \ac{ns-3} for the network simulation and the \ac{LIMoSim} kernel for mobility as proposed in \cite{Sliwa_etal_2019c} was selected due to its shared codebase approach, thereby allowing for a bidirectional influence between mobility and communication.

%
% Fig. Scenario
%
\fig{}{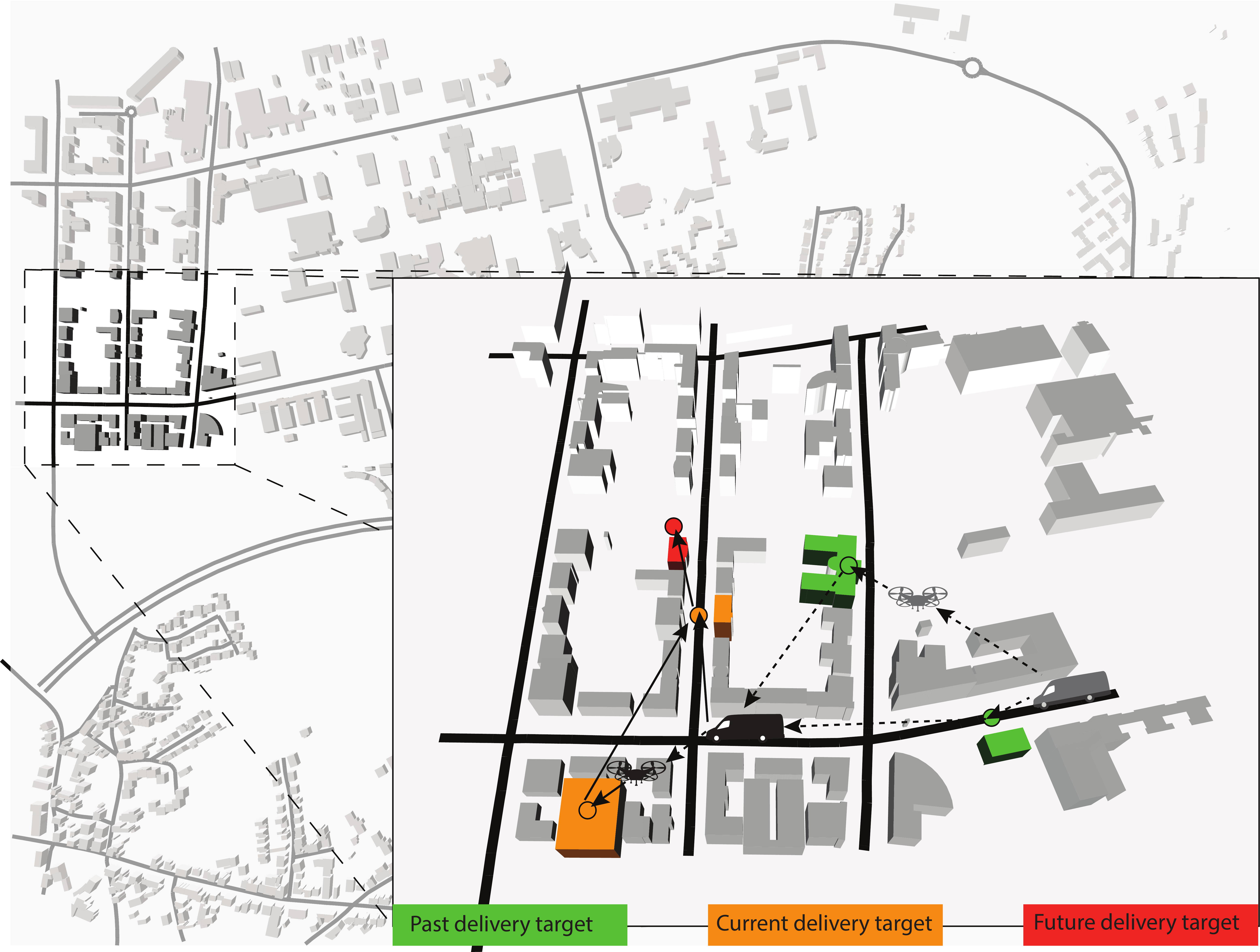}{Simulation scenario at the Dortmund technology center with an exemplary drone-aided parcel delivery illustration. }{fig:scenario}

The simulation scenario entails the buildings and road network around the Dortmund technology center and features a delivery truck and a specific number of \uavs.\\
%
% UAV mobility
%
The default \uav mobility model proposed by \ac{LIMoSim} is used for the delivery drones: a simplified version of the proposed model of \cite{Luukkonen/2011a}. With the angular vector $\eta = \begin{bmatrix} \theta \ \phi \ \psi \end{bmatrix}^{-1} $, which entails the values for \emph{pitch}, \emph{roll} and \emph{yaw}, motion in cartesian space is expressed as
\begin{equation}
\begin{bmatrix} \ddot{x} \\ \ddot{y} \\ \ddot{z} \end{bmatrix} =
-g \begin{bmatrix} 0 \\ 0 \\ 1 \end{bmatrix}
+ \frac{T}{m}
\begin{bmatrix} C_{\psi}S_{\theta}C_{\phi} + S_{\psi}S_{\phi} \\ S_{\psi}S_{\theta}C_{\phi} - C_{\psi}S_{\phi} \\ C_{\theta}C_{\phi} \end{bmatrix}
\end{equation}
where $C_{x}=\cos(x)$, $S_{x}=\sin(x)$, $g$ is the gravitation, $T$ is the thrust and $m$ the vehicle mass.
The angular accelerations are additionally modeled as
\begin{equation}
\mathbf{\ddot{\eta}} = \mathbf{J}^{-1}(\mathbf{\tau}_{B}-\mathbf{C}(\mathbf{\eta},\mathbf{\dot{\eta}})\mathbf{\dot{\eta}})
\end{equation}
where  $\mathbf{J}$ is a Jacobian matrix, which maps the angular velocities to angular accelerations, $\mathbf{\tau}_{B}$ being the torque and $\mathbf{C}(\mathbf{\eta},\mathbf{\dot{\eta}})$ being the Coriolis term, which accounts for gyroscopic and centripetal effects.

%
%	Truck mobility
%
For the delivery truck, a new vehicle model which extends the standard car model proposed by \ac{LIMoSim} with specialized parcel delivery navigation logic was developed.
Inheriting from the existing car model allows the delivery truck to rely on a hierarchical mobility model for considering different decision levels and control routines.

%
% Fig. Examplary delivery scenario
%
%\basicFig{}{fig/eps/delivery-plan}{Exemplary delivery scenario with the computed delivery plans}{fig:delivery_scenario}{0cm}{0cm}{0.8}

\begin{figure*}[]
	\centering	
	\includegraphics[width=1.8\columnwidth]{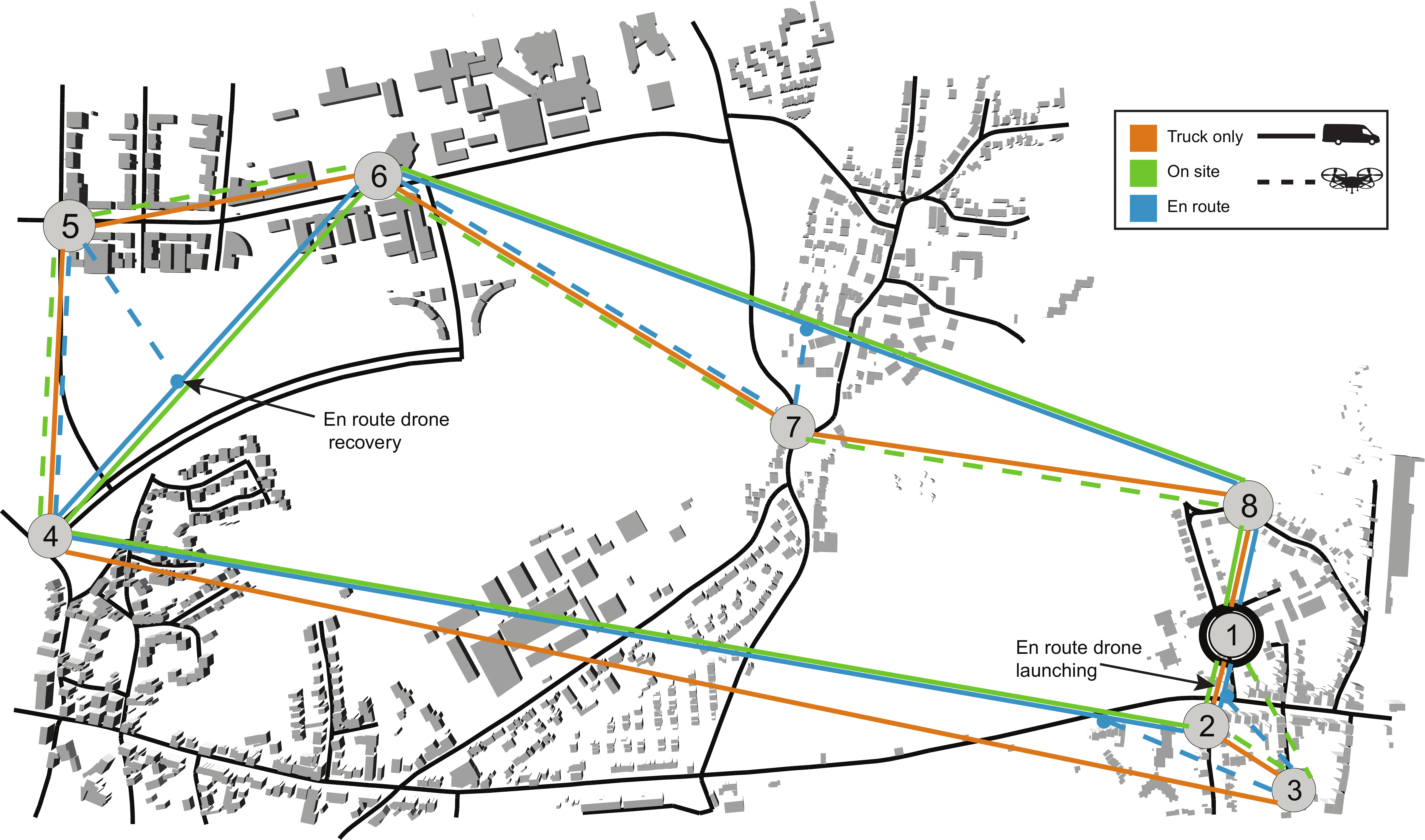}
	\caption{Exemplary delivery scenario with the computed delivery plans of the \truckonly, \onsite and \enroute approaches.}
	\label{fig:delivery_scenario}
\end{figure*}
%\dummyhere{Exemplary delivery Scenario}{Exemplary delivery scenario}{fig:delivery_scenario}
%
%
%
The truck must start from the depot center, visit each delivery location once and return to its start position.
A suitable route is therefore needed to complete the delivery as fast as possible.

%\subsection{Network Infrastructure}
%The scenario is adapted in different variants supporting the following network technologies to handle the communications between the truck and its drones ....

\subsection{\ac{TSP}-Modeling}

%
%	TSP
%
In order to generate the delivery route, the delivery scenario is treated as a \ac{TSP} and modeled as a graph \(G=(V,E)\) where the node \(v_0\) represents the depot and the nodes \((v_1,...,v_n)\) stand for the delivery recipients.
An optimal solution must satisfy the following constraint:
\begin{equation}
min\sum_{i=1}^{n}\sum_{j\neq i,j=1}^{n}c_{ij} \cdot x_{ij}
\end{equation}
with the cost \(c_{ij}\) being the distance between \(v_i\) and \(v_j\) and \(x_{ij}\) a path selection variable obeying:
\begin{equation}
x_{ij}=
\begin{cases}
1       & \quad \text{if path goes from } v_i \text{ to } v_j\\
0		& \quad \text{otherwise}
\end{cases}
\end{equation}

The following additional constraints must apply to an optimal solution:

\begin{equation}
\label{eqn:solution_constraint_1}
\forall j \in (1,...,n) \sum_{i=1, i\neq j}^{n} x_{ij} = 1
\end{equation}

The constraint in Eq.~\ref{eqn:solution_constraint_1} enforces that each node is visited exactly once.
%
% Graph, Costs, Routing
%
The cost of each edge \(c_{ij}\) is determined by taking the shortest route between the edge's nodes, based on the \ac{OSM} road network data which is loaded in \ac{LIMoSim} \cite{Sliwa_etal_2019c}.
While searching for the shortest route between two nodes, u-turns are only allowed at dead ends or just after completing a delivery.\\
%
% TSP solving
%
The Lin-Kernighan heuristic is used to solve the thus modeled \ac{TSP} instance, as it produces nearly optimal solutions for large \ac{TSP} instances with less computational effort than the optimal integer programming method \cite{Rego2011}.
%
% TSP usage
%
The \ac{TSP} solution is then transformed into a \ac{LIMoSim} route for the delivery truck consisting of a sequence of gates, which must be taken to leave each encountered intersection.\\
%\basicFig{}{fig/eps/example-delivery-truck}{Truck route for the example scenario without \uav-enhancement}{fig:tsp_example}{0cm}{0cm}{0.8}

\subsection{Enhancement through Drone Delivery}
%
% 	TSP-D with variants
%
The \ac{TSP} solution is then enhanced to account for drone support in two different variants, thus producing two \ac{TSP-D} tours.\\
%
% On-site
%
The first variant is \onsite drone dispatching and recovery, which means the drones are only dispatched and recovered at truck delivery nodes.\\
%
% En-route
%
The second variant is \enroute drone dispatching and recovery, which as opposed to the first variant, allows for drones to be deployed and recovered at any location on the truck's route. this variant may yield delivery efficiency gains over the on-site variant, but can at the same time be difficult or not always possible to perform. For instance, recovering a drone en route while the truck is riding on a highway may prove to be a tedious landing task.\\

\begin{figure}[]
	\centering	
	\subfig{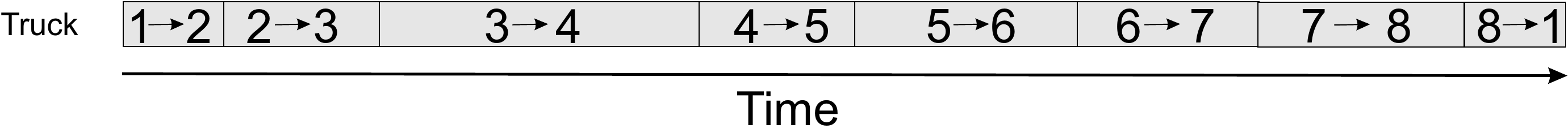}{0.5}{Truck only}
	\subfig{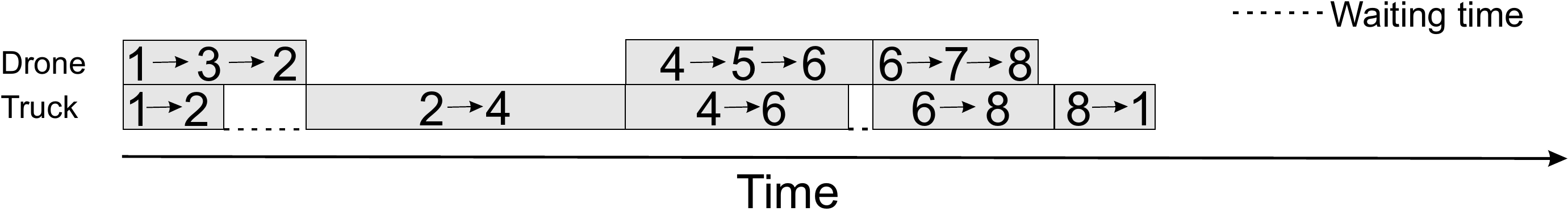}{0.5}{On site}
	\subfig{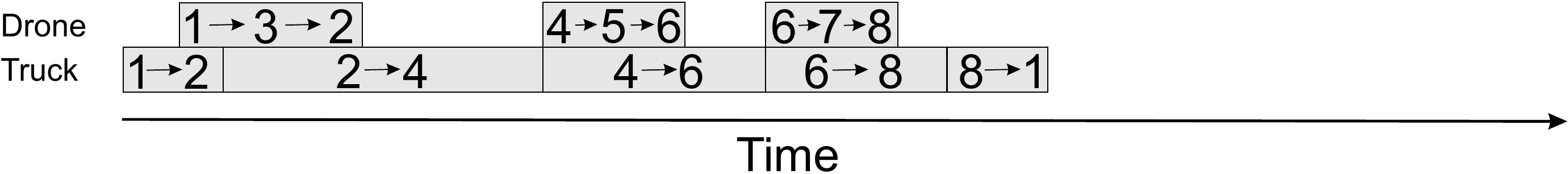}{0.5}{En route}
	\caption{Gantt charts of the different delivery methods applied on the example scenario.}
	\label{fig:gantt_tours}
\end{figure}

%
% Drone job attribution scheme
%
A basic job scheduling scheme is used to determine which delivery jobs are assigned to \uavs. A job gets allocated to a \uav if it is available and when the total estimated travel time for the job is smaller than the maximal allowed travel time for a drone.
As a consequence, the job scheduling scheme does not disrupt the truck nodes order as it simply removes jobs from the truck's list and assigns them to \uavs.
The truck route is then constructed from the \ac{TSP-D} solution analog to the \ac{TSP} case.
The generated solutions for the TSP, \ac{TSP-D} and \ac{TSP-D} with en route operations for our example scenario are illustrated in Fig.~\ref{fig:delivery_scenario} and the associated Gantt charts are presented in Fig.~\ref{fig:gantt_tours}.

%\begin{figure*}[]
%	\centering	
%	\subfig{fig/eps/example-delivery-uav-onsite}{0.4}{On site}
%	\subfig{fig/eps/example-delivery-uav-enroute}{0.4}{En route}	
%	\caption{Delivery routing for the truck and the \uavs for the example scenario}
%	\label{fig:tsp-d_example}
%\end{figure*}

%\fig{}{fig/eps/example-delivery}{Truck route for the example scenario without \uav-enhancement}{fig:tsp_example}
%\dummyhere{TSP-D solution on-site and en route}{Delivery routing for the truck and the \uavs for the example scenario}{fig:tsp-d_example}
%\dummyhere{En route enhancement}{\uav-enhanced delivery routing with en route dispatching and recovery}{fig:tsp-d_en_route_example}

%
% Runtime parameters
%
To further mimic the logistics of delivery, execution parameters are used to control the waiting time for unloading the parcel at a delivery target once the agent - the truck or a drone - reaches it (further referred to as unloading time), as well as the take-off and landing duration of the drones. Speed limitations on the different road segments are respected. However, traffic signaling and congestion were not considered.\\

%
% TSP-D Protocol
%
To fully leverage the potential of a heterogeneous delivery fleet requires a means to synchronize the agents.
To this end, we defined a protocol to coordinate the interactions between the delivery truck and the \uavs, which is graphically featured in Fig.~\ref{fig:protocol}.

\fig{}{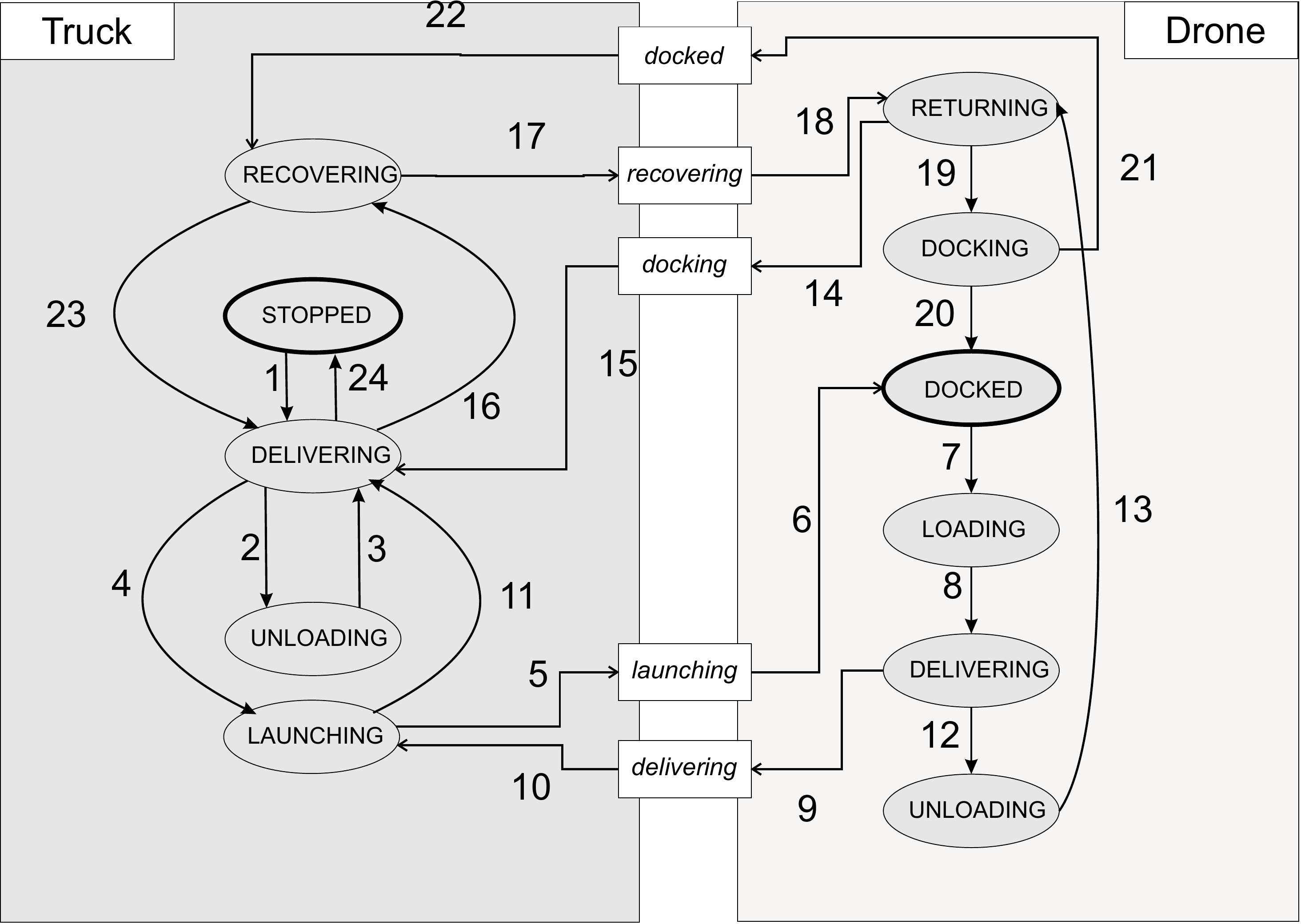}{Interaction protocol used to coordinate the heterogeneous fleet of delivery agents.}{fig:protocol}
%\dummyhere{Protocol Chart}{Interaction protocol of delivery agents}{fig:protocol}

The delivery truck may deliberately wander between the \texttt{STOPPED}, \texttt{DELIVERING}, \texttt{UNLOADING} and \texttt{LAUNCHING} states.\\
For instance, upon reaching a delivery target, the truck switches to the \texttt{UNLOADING} state to unload the parcel and complete delivery.
The truck then switches back to the \texttt{DELIVERING} state after the unloading time is up and resumes delivery.\\
Nearly the same cycle is observed for the \texttt{LAUNCHING} state on the truck side with the only exception that there is no launch timer.
The delivery truck switches to the \texttt{LAUNCHING} state when it reaches a drone launch point and leaves it upon receiving a launch completion message from the launched \uav.\\
Entering and leaving the \texttt{RECOVERING} state is triggered by any \uav aiming to regroup with the delivery truck.
Before recovery, the truck slows down upon sensing a returning drone within a specific range and stops once the drone initiates recovery.

The drones get started in the \texttt{DOCKED} state, where they are docked to the truck and can, therefore, travel through the scenario without exerting any effort.
Upon receiving the launch message from the truck, they change in the \texttt{LOADING} mode, which they then leave for the \texttt{DELIVERING} state after the take-off timeout and notify the truck of launch completion.\\
Once they reach the delivery target, they switch to \texttt{UNLOADING} and continue in the \texttt{RETURNING} state once the unloading duration is spent.\\
While returning, the drones send a docking message to the truck as they get in docking range, causing the truck to stop for them to complete the docking maneuver after which they notify the truck of docking completion.\\

The en route delivery variant requires additional logic to keep the truck waiting in optimal launching or recovery range until the awaited events occur.
In the on-site delivery variant, every truck delivery target is a potential drone launch or recovery point.

%\dummyhere{Gantt charts of the tours}{Gantt charts of each of the tour variants for the example scenario}{fig:gantt_tours}.

\subsection{Energy Consumption}
%
% Battery
%
To assess the power efficiency of the various delivery methods, the internal energy monitoring of \ac{LIMoSim} is used.
The mobility-related energy consumption tracking in \ac{LIMoSim} is enabled through its hierarchical mobility model.
The control inputs which are sent to the mobility's locomotion layer are forwarded fo to a numerical energy model which computes the resulting consumption, based on the time interval during which these inputs are applied on the locomotion.

The energy model used tracks only the mobility-related consumption of \uavs.
Further details to the energy model have been published in \cite{Sliwa_etal_2019c}.

\subsection{Mobility Prediction}
%
% Prediction
%
To further optimize the energy footprint of the drone delivery, we consider mobility prediction as it is one of the key enabling methods for anticipatory communication \cite{Sliwa/etal/2019a}.

\ac{LIMoSim} supports mobility prediction and allows the usage of a customized model.
However it  inherently provides an implementation of a hierarchical model which iteratively uses different prediction methods for forecasting successive future vehicle positions. At each prediction iteration, the method with the highest expected precision is used as illustrated in Fig.~\ref{fig:mobility_prediction}.

\fig{}{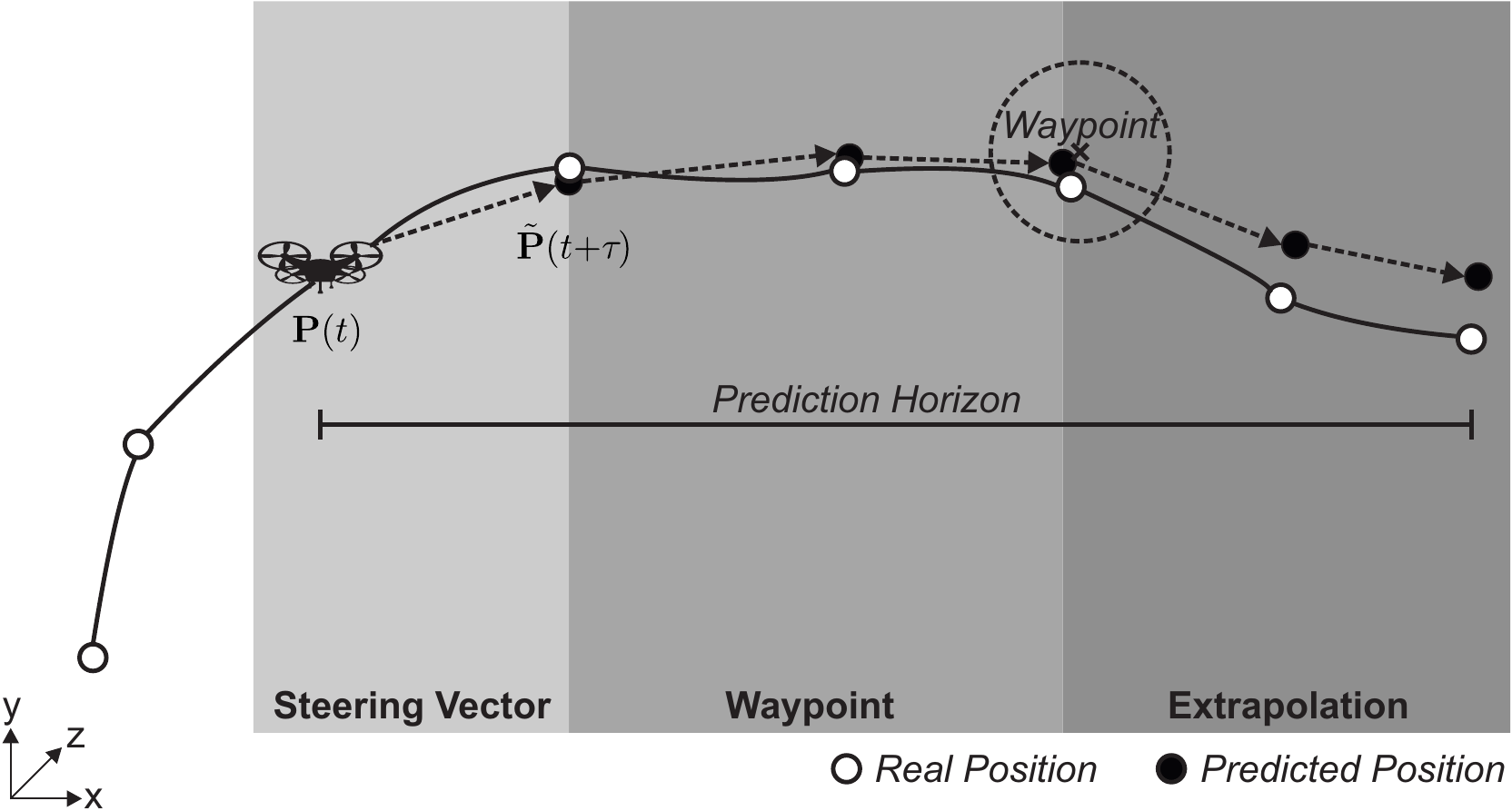}{Hierarchical mobility prediction model overview according to \cite{Sliwa_etal_2019c}. The method with the highest prediction accuracy is selected in each iteration step.}{fig:mobility_prediction}

Specific extensions were made to the provided mobility prediction to leverage the routing information contained in the delivery route depending on the drone deployment scheme. These extensions are the following:
\begin{itemize}
	\item In on site mode, the next waypoint is the delivery target where regrouping shall take place
	\item In en route mode, the next waypoint may be the next road intersection on the truck's route or the next delivery target
\end{itemize}

%\section{Methodology} \label{sec:methods}

%
% Fig. Screenshot
%
%\dummy{Screenshot}{arg2}{fig:map}
%
%
%
%Fig.~\ref{fig:map}

%
% Tab. Parameters
%
%\input{tex/tables/parameters.tex}
%
%
%
%Tab.~\ref{tab:parameters}
\section{Evaluation of the Delivery Approaches}
\label{sec:results}

In this section, we evaluate the performance of different delivery approaches.
\ac{LTE} and \ac{C-V2X} (Mode 4) are applied for communications among delivery agents.
The \ac{C-V2X} implementation developed in \cite{Eckermann/etal/2019a} is used.

%
% Tab. General Evaluation Parameters
%
\newcommand{\entry}[2]{#1 & #2 \\}
\newcommand{\layeredEntry}[2]{& #1 & #2 \\}
\newcommand{\head}[2]{\toprule \entry{\textbf{#1}}{\textbf{#2}}\midrule}

\newcommand{\sideHeader}[3]
{
	\multirow{#1}{*}{
		\rotatebox[origin=c]{90}{
			\parbox{#2}{\centering \textbf{#3}}
		}
	}
}

\begin{table}[ht]
	\centering
	\caption{General simulation parameters}
	\begin{tabular}{ll}
		%		\sideHeader{13}{0.5cm}{General} 
		\head{Parameter}{Value}
		%		\entry{\textbf{Parameter}}{\textbf{Value}}
		
		\entry{Scenario size}{3000~m x 1500~m x 250~m}
		
		\entry{Maximum speed (Truck)}{13~$m/s$}
		\entry{Maximum speed (UAV)}{16~$m/s$}
	
		\bottomrule
		
	\end{tabular}
	\label{tab:general-parameters}
\end{table}

The general simulation parameters common to all evaluations presented hereafter are listed in Tab.~\ref{tab:general-parameters}

\subsection{Enhancement of Delivery Performance}

%
%	TSP vs TSP-D vs TSP-D en route
%
In the first analysis, three delivery modes are analyzed: \truckonly, \onsite and \enroute. In the drone assisted modes, we study the effects of increased drone usage.
A summary of the relevant simulation parameters of this evaluation is given in Tab.~\ref{tab:mobility-evaluation-parameters}.

%
% Tab. Mobility Evaluation Parameters
%
%\newcommand{\entry}[2]{#1 & #2 \\}
%\newcommand{\head}[2]{\toprule \entry{\textbf{#1}}{\textbf{#2}}\midrule}
%
%\newcommand{\sideHeader}[3]
%{
%	\multirow{#1}{*}{
%		\rotatebox[origin=c]{90}{
%			\parbox{#2}{\centering \textbf{#3}}
%		}
%	}
%}

\begin{table}[ht]
	\centering
	\caption{Parameters of the logistics enhancement evaluation}
	\begin{tabular}{ll}
%		\sideHeader{13}{0.5cm}{General} 
		\head{Parameter}{Value}
%		\entry{\textbf{Parameter}}{\textbf{Value}}
		
		\entry{Simulation duration}{until delivery completion}
		\entry{Delivery sets}{50}
		\entry{Deliveries per set}{15}
		\entry{UAV count}{\{0,~...,~15\}}

		\bottomrule
		
	\end{tabular}
	\label{tab:mobility-evaluation-parameters}
\end{table}

\fig{!h}{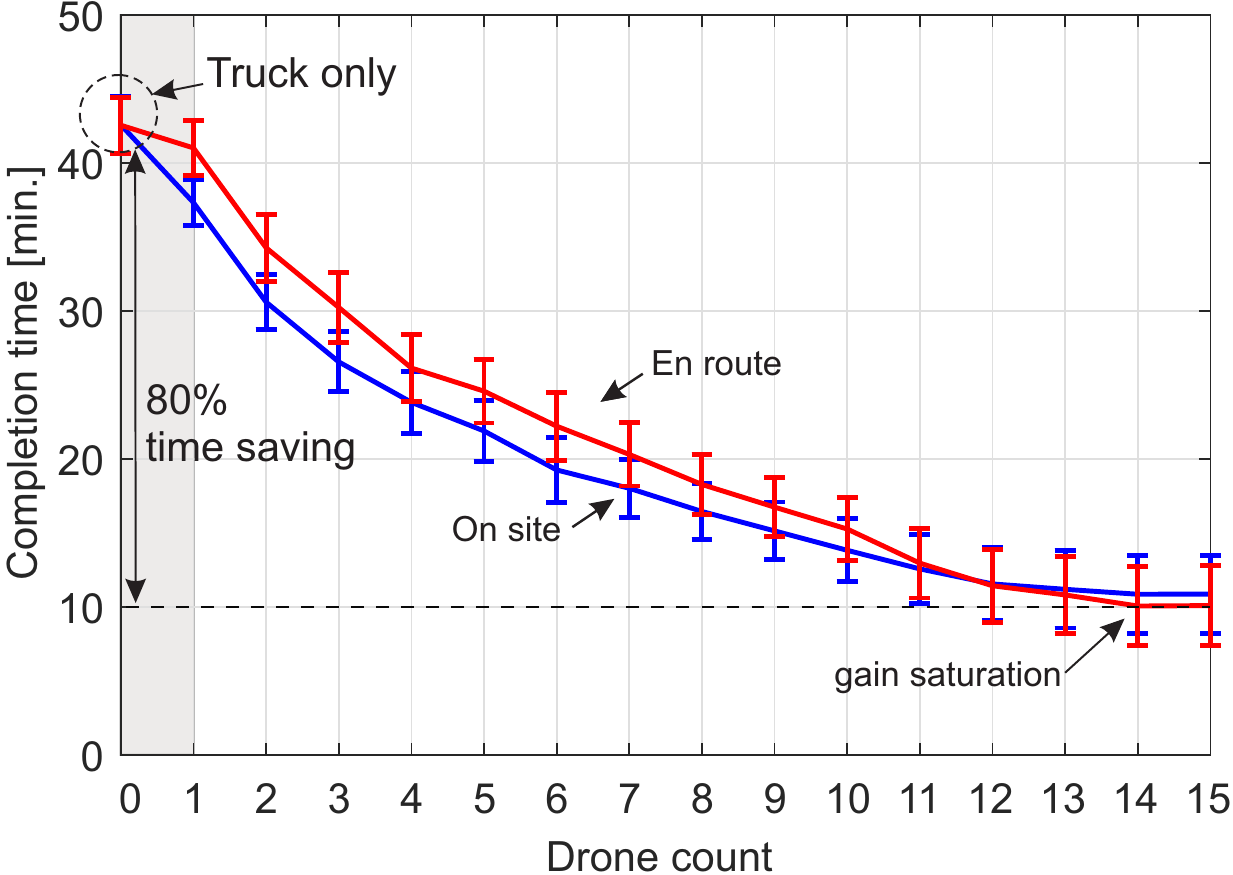}{Completion time of delivery using the truck only mode as well as the on site and en route approaches with an increasing number of drones.}{fig:completion-time-more-uavs}

The evolution of the completion time with an increasing number of deliverer drones in Fig.~\ref{fig:completion-time-more-uavs} shows that the delivery tour can be completed faster with the usage of more drones.
However, the time gain decreases after each drone increase until saturation is reached.
It can also be observed that switching the operations mode from \onsite to \enroute adds a supplementary delay on the overall completion time.
This is due to the simplicity of the drone delivery scheduling scheme used. As mentioned in Sec.~\ref{sec:approach}, the used scheme assigns deliveries to the drones while keeping the order of the remaining truck deliveries consistent with the truck only delivery schedule, therefore, causing the truck to stop more often in en route mode.
It was shown that the en route approach yields better completion times than on site \cite{Marinelli2018}, however, a more complex job scheduling scheme must be used to further leverage the mobility advantage gained from drone support.

\subsection{Energy Savings and Mobility Prediction}

%
% Problemativ
%
In the second evaluation, we focus on the energy consumption of the delivery modes which were used in the previous evaluation. Furthermore, we investigate the impact of mobility prediction on energy consumption in en route drone operations.
This is motivated by the fact that drone recovery, which requires a regrouping point with the truck, does not occur at fixed points in en route mode as opposed to on site mode.
% and is most likely subject to changes depending on the truck's mobility.

%
% Tab. Energy Evaluation Parameters
%
%\input{tex/tables/energy-evaluation-parameters.tex}
%
%
%

%\fig{!h}{results/fig/delivery_stats_energy_uavcount}{}{fig:en-route-savings-uavcount}
%\fig{!h}{results/fig/delivery_stats_energy_mode}{Completion time of drone with en route operations with increasing uav count}{fig:en-route-savings-mode}

\fig{!h}{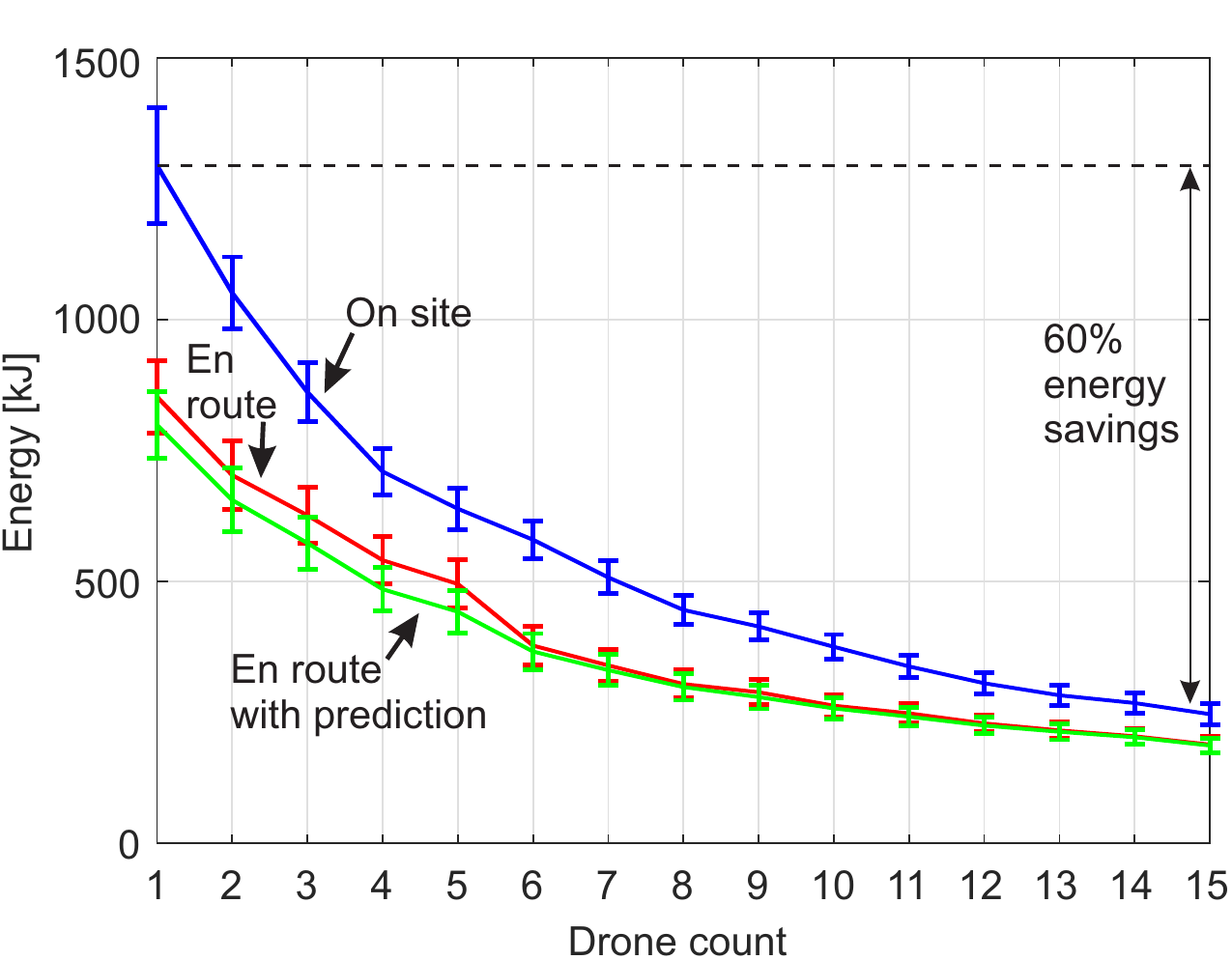}{Energy consumption per drone with increasing drone count for the different delivery approaches.}{fig:energy-uav-count}

\begin{figure}[]
	\centering	
	\subfig{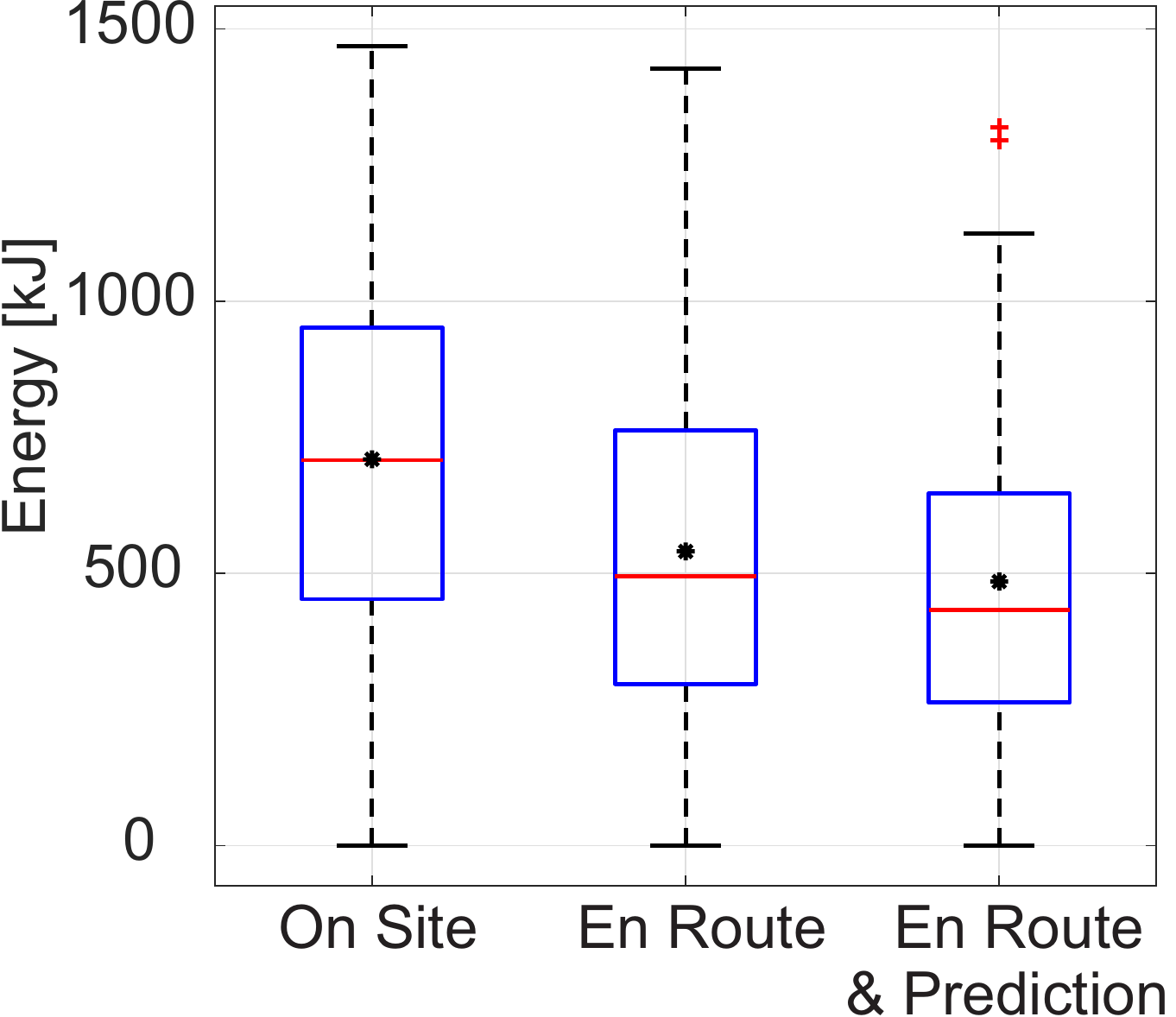}{0.22}{4 drones}
	\subfig{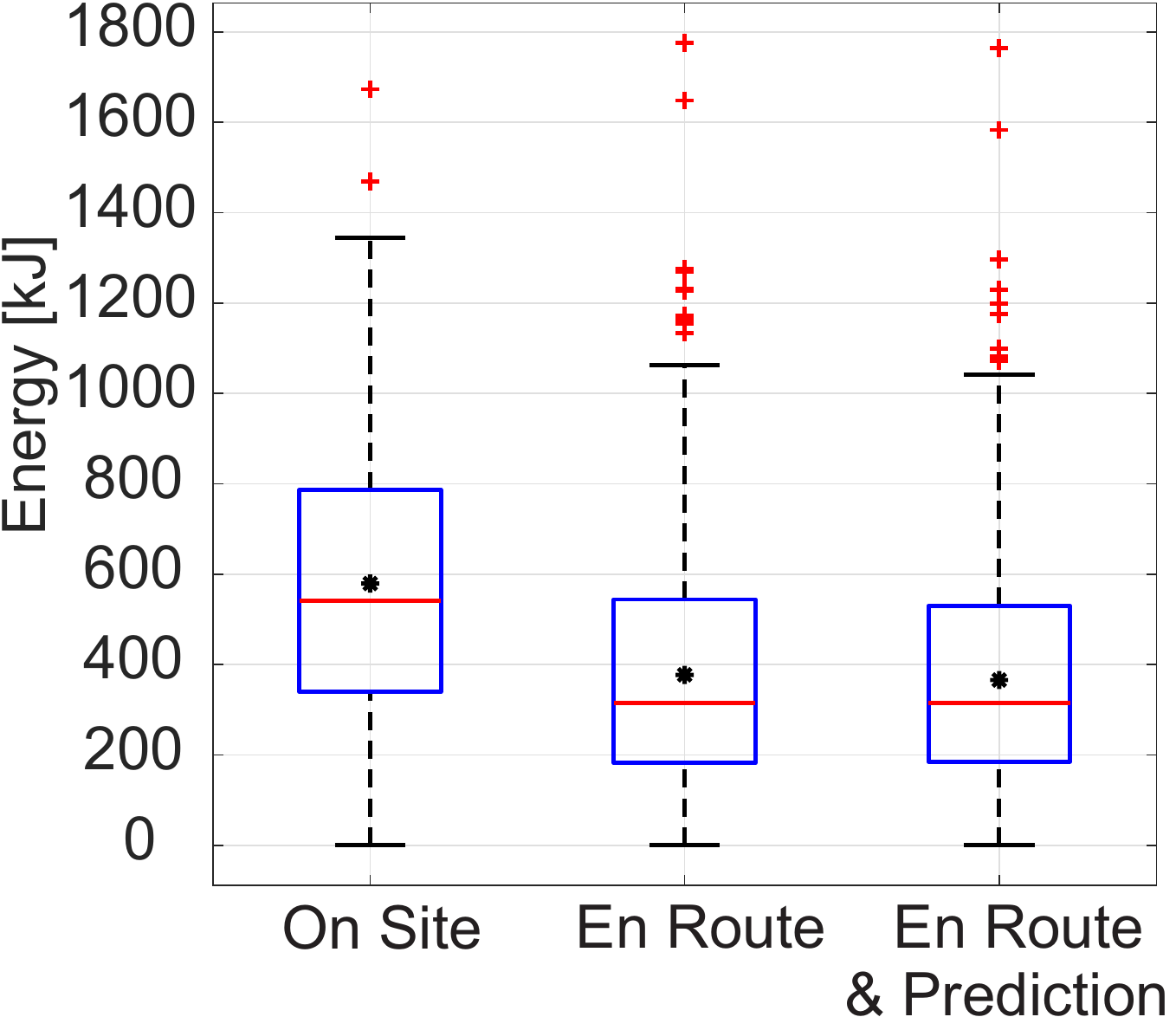}{0.22}{6 drones}	
	\caption{Energy consumption per drone of the different delivery approaches, before and after the efficiency gain of mobility prediction wears off}
	\label{fig:en-route-savings}
\end{figure}

The mean energy consumption per \uav and grouped per delivery mode is shown in Fig.~\ref{fig:energy-uav-count}.
An overall reduction of the energy footprint per drone is achieved when switching from on site to en route operations.
Using more drones can also decrease the average energy consumption per drone up to 60~\%. 
Mobility prediction allows for further energy sparing, but its effects become less perceptible as the drone count increases.
This is presented more explicitly in Fig.~\ref{fig:en-route-savings} for the evaluated scenario. When using less than 6 drones, for instance 4, the sparing effects of mobility prediction are noticeable.
The energy saved through mobility prediction becomes negligible from 6 drones upwards.

\subsection{Investigating Network Communication Standard Usability}

In the second evaluation, we monitor mission-critical network metrics for \uav-empowered \ac{PDP} applications: the transmission latency and the \ac{PDR}.
The criticalness of these metrics, which is common among robotics applications, arises from the requirement of permanent monitoring of the \uavs.
Therefore a reliable communication link between the delivery drones and the truck is needed.

%
%	LTE vs C-V2X
%
To carry out our evaluation, standard \acp{CAM} are sent from the \uavs to the truck with a size of 190~Bytes every 100~ms \cite{Eckermann/etal/2019a}. These messages are likely safety-related, so the delay must be kept low and \ac{PDR} should be high.
Two communication technologies are applied: \ac{LTE} and \ac{C-V2X}. Common operating parameters values are used for the network configuration.
An overview of the relevant simulation parameters to this evaluation is given in Tab.~\ref{tab:communication-evaluation-parameters}.

%Simulative results on how each communication standard fares with respect specific metrics (delay, PDR):
%\begin{itemize}
%	\item LTE
%	\item C-V2X
%\end{itemize}
%
% Tab. Communication Evaluation Parameters
%
%\newcommand{\layeredEntry}[2]{& #1 & #2 \\}
%\newcommand{\head}[2]{\toprule \entry{\textbf{#1}}{\textbf{#2}}\midrule}
%
%\newcommand{\sideHeader}[3]
%{
%	\multirow{#1}{*}{
%		\rotatebox[origin=c]{90}{
%			\parbox{#2}{\centering \textbf{#3}}
%		}
%	}
%}

\begin{table}[ht]
	\centering
	\caption{Parameters of the communication technology evaluation}
	\begin{tabular}{p{0.2cm}p{2.8cm}p{4.5cm}}
		\toprule
		\sideHeader{8}{3cm}{General} 
%		\head{Parameter}{Value}
		\layeredEntry{\textbf{Parameter}}{\textbf{Value}}
		\midrule
		
		\layeredEntry{Simulation duration}{15~min}
		\layeredEntry{Delivery sets}{50}		
		\layeredEntry{Deliveries per set}{15}
		\layeredEntry{UAV count}{\{~1,~2,~3,~4,~5~\}}		
		\layeredEntry{Channel model}{ DeterministicObstacleShadowing}
		
		\layeredEntry{Packet Size}{190~Byte}
		\layeredEntry{Inter Packet Interval}{100~ms}
		
		%
		% V2X
		%
		\midrule
		\sideHeader{4.5}{0.3cm}{\mbox{C-V2X}} 
		\layeredEntry{Carrier frequency}{5.9~GHz}
		\layeredEntry{Bandwidth}{20~MHz}
		\layeredEntry{$P_{\text{TX}}$ (\acs{UE})}{23~dBm}
		
		%
		% LTE Carrier-Frequency
		%
		\midrule
		\sideHeader{5}{0.3cm}{LTE} 
		\layeredEntry{Carrier frequency}{2.1~GHz}
		\layeredEntry{Bandwidth}{20~MHz}
		\layeredEntry{$P_{\text{TX}}$ (\acs{UE})}{23~dBm}
		\layeredEntry{$P_{\text{TX}}$ (\acs{eNB})}{43~dBm}

		\bottomrule
		
	\end{tabular}
	\vspace{0.1cm}
	$P_{\text{TX}}$: Maximum transmission power
	\label{tab:communication-evaluation-parameters}
\end{table}

%
%
%

%\fig{!h}{results/fig/lte_vs_cv2x_delay}{Completion time of drone with en route operations with increasing uav count}{fig:completion-time-more-uavs}

\begin{figure}[h!]
	\centering	
	\subfig{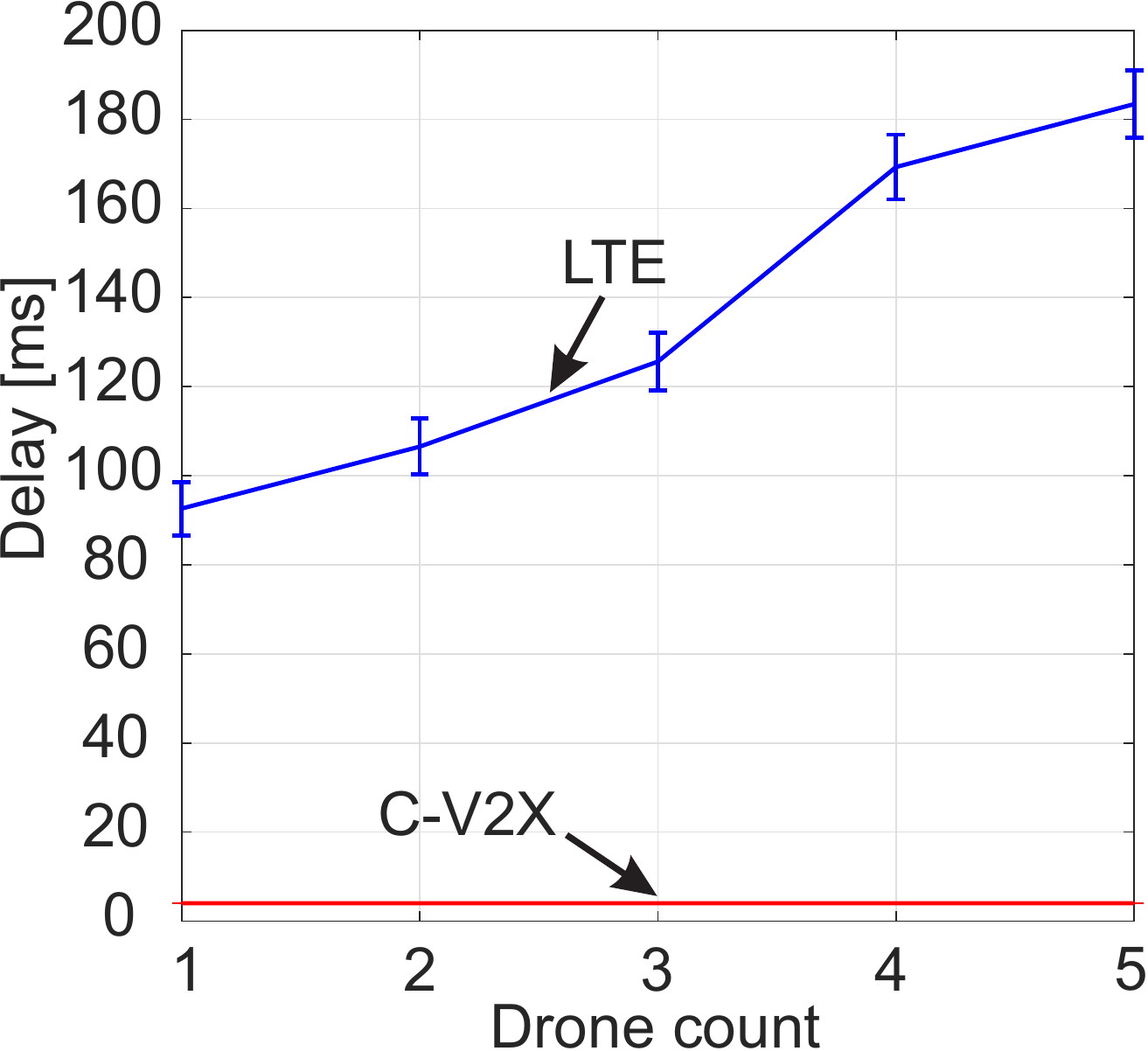}{0.22}{}
	\subfig{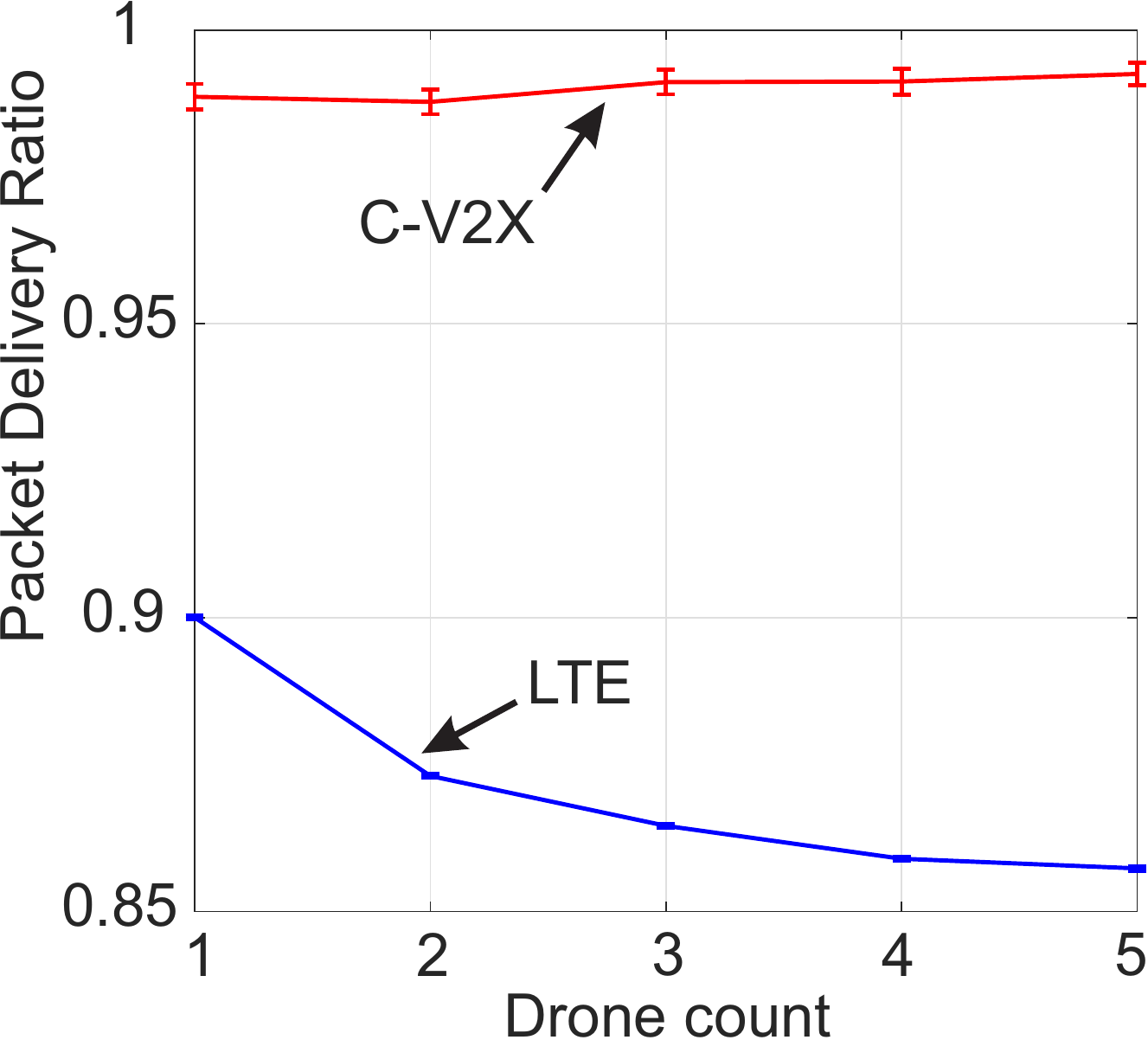}{0.22}{}	
	\caption{Comparison between \ac{LTE} and \ac{C-V2X} for connecting the delivery drones to the truck.}
	\label{fig:communication-evaluation}
\end{figure}

The statistical evaluation of the results is illustrated in Fig.\ref{fig:communication-evaluation} and shows that \ac{C-V2X} achieves a significant smaller latency than \ac{LTE}. This stems from the applied \ac{C-V2X} variant which implements \textit{Mode 4} behavior \cite{Eckermann/etal/2019a}. The channel is accessed directly, which gives the intra-\ac{UE} processing the dominant share on the overall delay.
Whereas \ac{LTE} transmissions are subject to resource scheduling mechanisms of the \ac{eNB}.
The direct channel access of \ac{C-V2X} however increases the
collision probability with a growing number of \ac{UE}s operating in the same interference region but does not hinder \ac{C-V2X} from achieving a higher \ac{PDR} than \ac{LTE} with an increased number of drones as \acp{UE}. The \ac{LTE} link quality is affected by the shadowing of buildings present in the scenario. These have an average height of 20~m.
%thus leading to reduced \ac{PDR}.

%\dummyhere{\ac{LTE} vs \ac{C-V2X}}{Network latency and \ac{PDR} during truck delivery with increasing drone count}{fig:communication-evaluation}

\subsection{The Influence of Building's Height}

In this last evaluation, we qualitatively investigate the influence of buildings height on the \ac{LTE} communication link in the parcel delivery application context.
The \ac{LTE} simulation setup of the previous evaluation is used and the mean of the building's height distribution in the scenario is varied.
%The simulation setup of the previous evaluation is used.

In order to model a deterministic and consistent interference behavior of the buildings the \textbf{deterministic obstacle shadowing model} provided in \ac{LIMoSim} is used.
An exemplary \ac{RSRP} heatmap of the simulation scenario generated using this model is shown in Fig.\ref{fig:heatmap}.

\basicFig{!h}{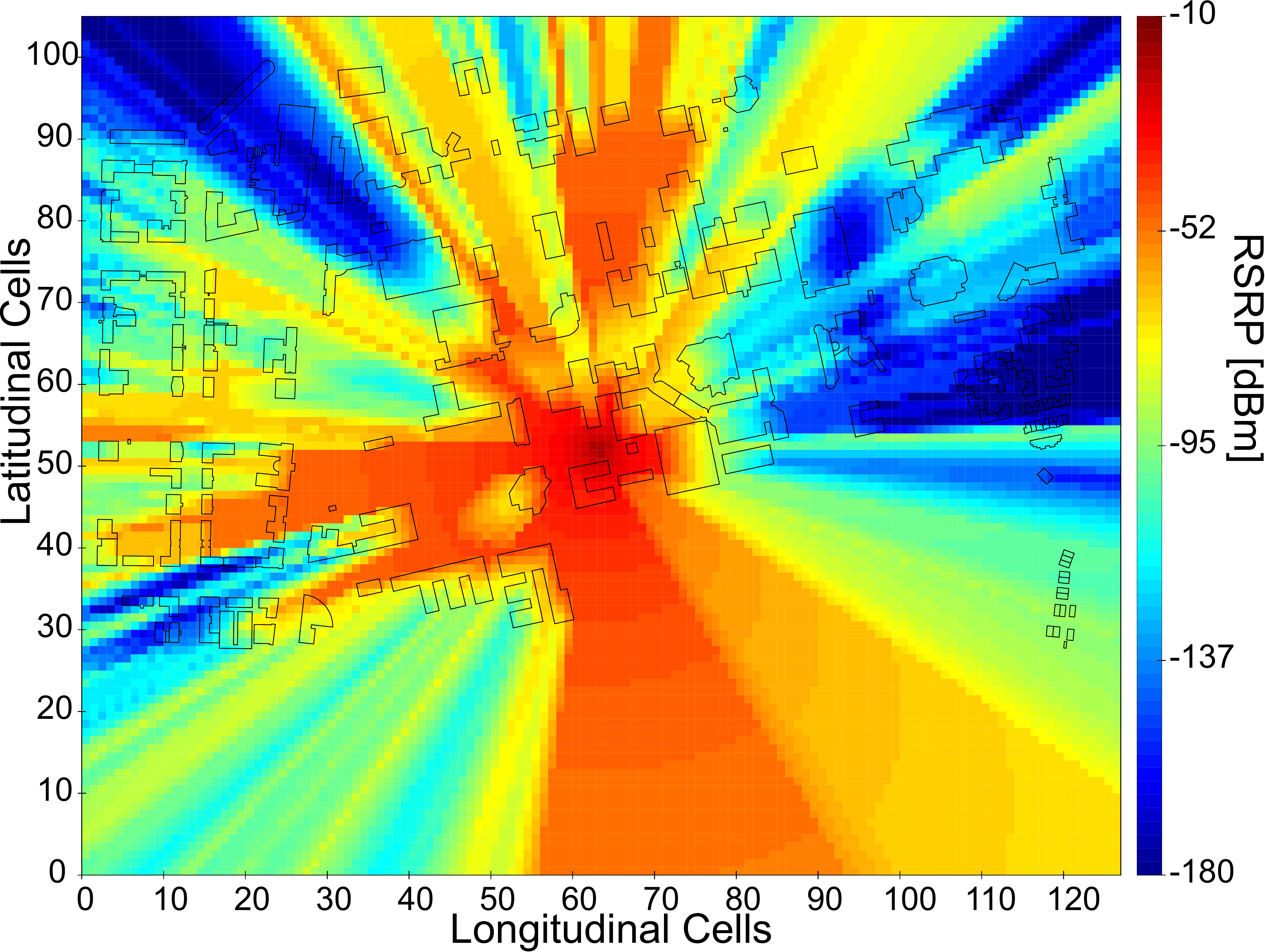}{Exemplary signal strength map at 10~m above ground derived by the obstacle shadowing model.}{fig:heatmap}{0cm}{0cm}{0.8}

%
% Tab. Building influence Evaluation Parameters
%
%\newcommand{\layeredEntry}[2]{& #1 & #2 \\}
%\newcommand{\head}[2]{\toprule \entry{\textbf{#1}}{\textbf{#2}}\midrule}
%
%\newcommand{\sideHeader}[3]
%{
%	\multirow{#1}{*}{
%		\rotatebox[origin=c]{90}{
%			\parbox{#2}{\centering \textbf{#3}}
%		}
%	}
%}

\begin{table}[ht]
	\centering
	\caption{Parameters of the buildings influence evaluation}
	\begin{tabular}{p{0.2cm}p{2.8cm}p{4.5cm}}
		\toprule
%		\sideHeader{13}{0.5cm}{General} 
%		\head{Parameter}{Value}
		\layeredEntry{\textbf{Parameter}}{\textbf{Value}}
		\midrule	
		\layeredEntry{Simulation duration}{15~min}
		\layeredEntry{Delivery sets}{50}		
		\layeredEntry{Deliveries per set}{15}
		\layeredEntry{UAV count}{\{~1,~2,~3,~4,~5~\}}		
		\layeredEntry{Channel model}{ DeterministicObstacleShadowing}
		\layeredEntry{Mean building height}{\{~10~m,~20~m,~30~m\}}
	
		\bottomrule
		
	\end{tabular}
	\vspace{0.1cm}
	\label{tab:building-evaluation-parameters}
\end{table}

\begin{figure}[h!]
	\centering	
	\subfig{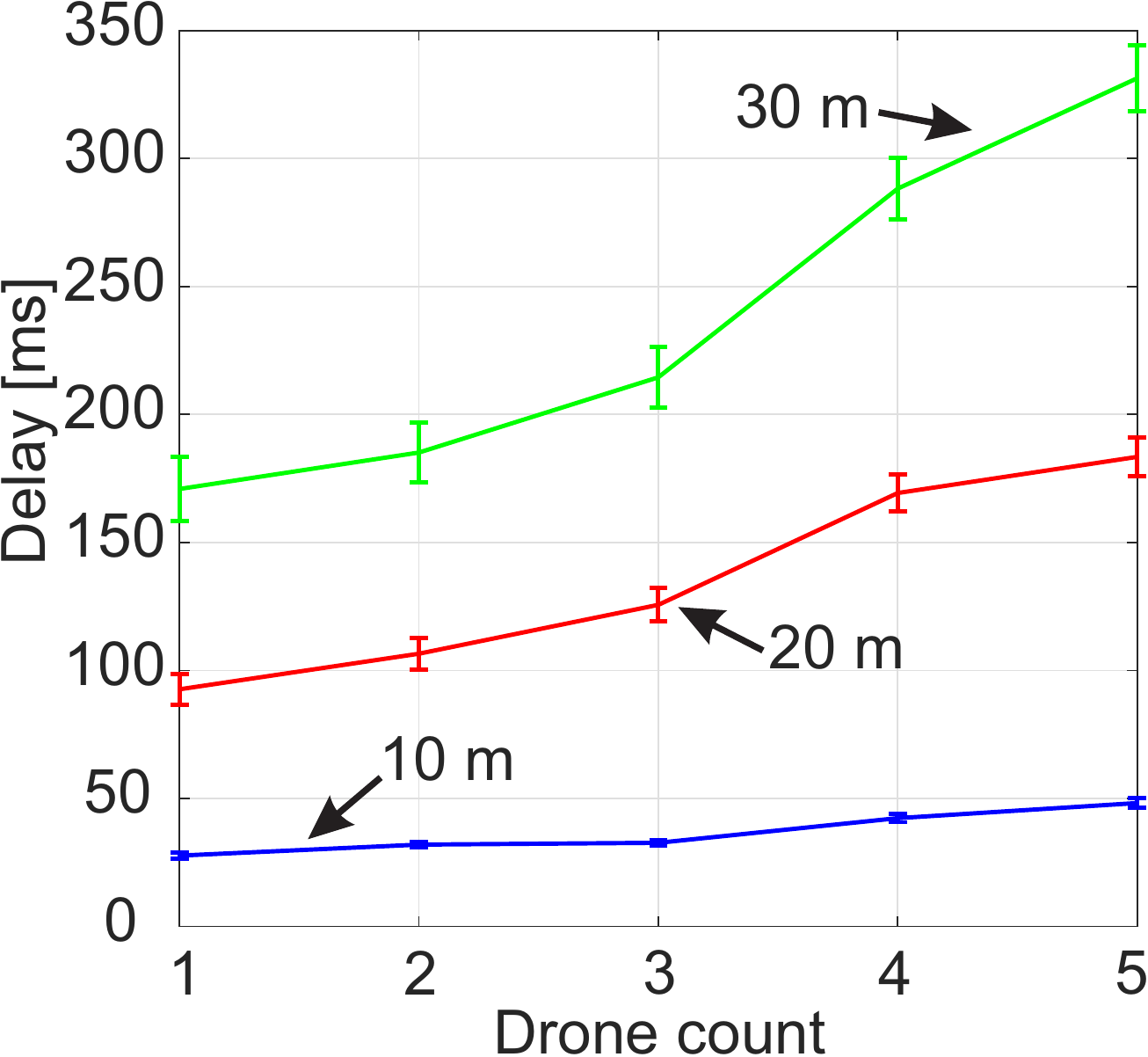}{0.22}{}
	\subfig{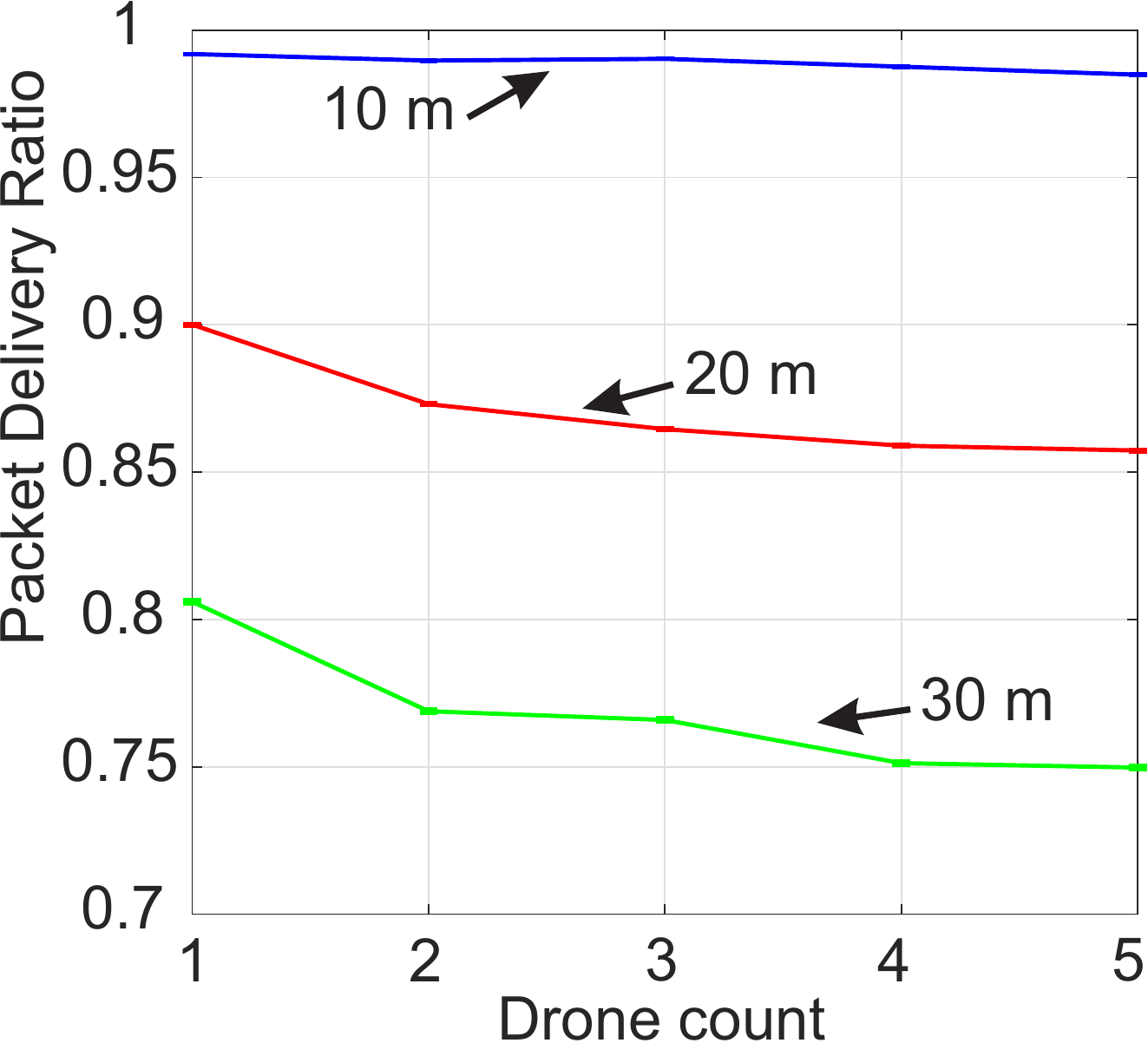}{0.22}{}	
	\caption{Performance of the \ac{LTE} link connecting drones to the truck with  different mean height of buildings.}
	\label{fig:communication-evaluation-height}
\end{figure}

The increasing heights of buildings is shown in Fig.~\ref{fig:communication-evaluation-height} to have a reinforcing effect on the network latency.
This is due to the poor radio channel conditions associated with the shadowing caused  by tall buildings.
A similar effect befalls the \ac{PDR}, which gets smaller with greater buildings height.
Furthermore, having more drones access the channel increases the delay on a greater scale when surrounded by high buildings. The results show that the delay's growth rate is higher when increasing the drone count at bigger buildings height. The \ac{PDR} also degrades faster with an increasing number of \acp{UE} around high buildings.

\begin{figure}[h!]
	\centering	
	\subfig{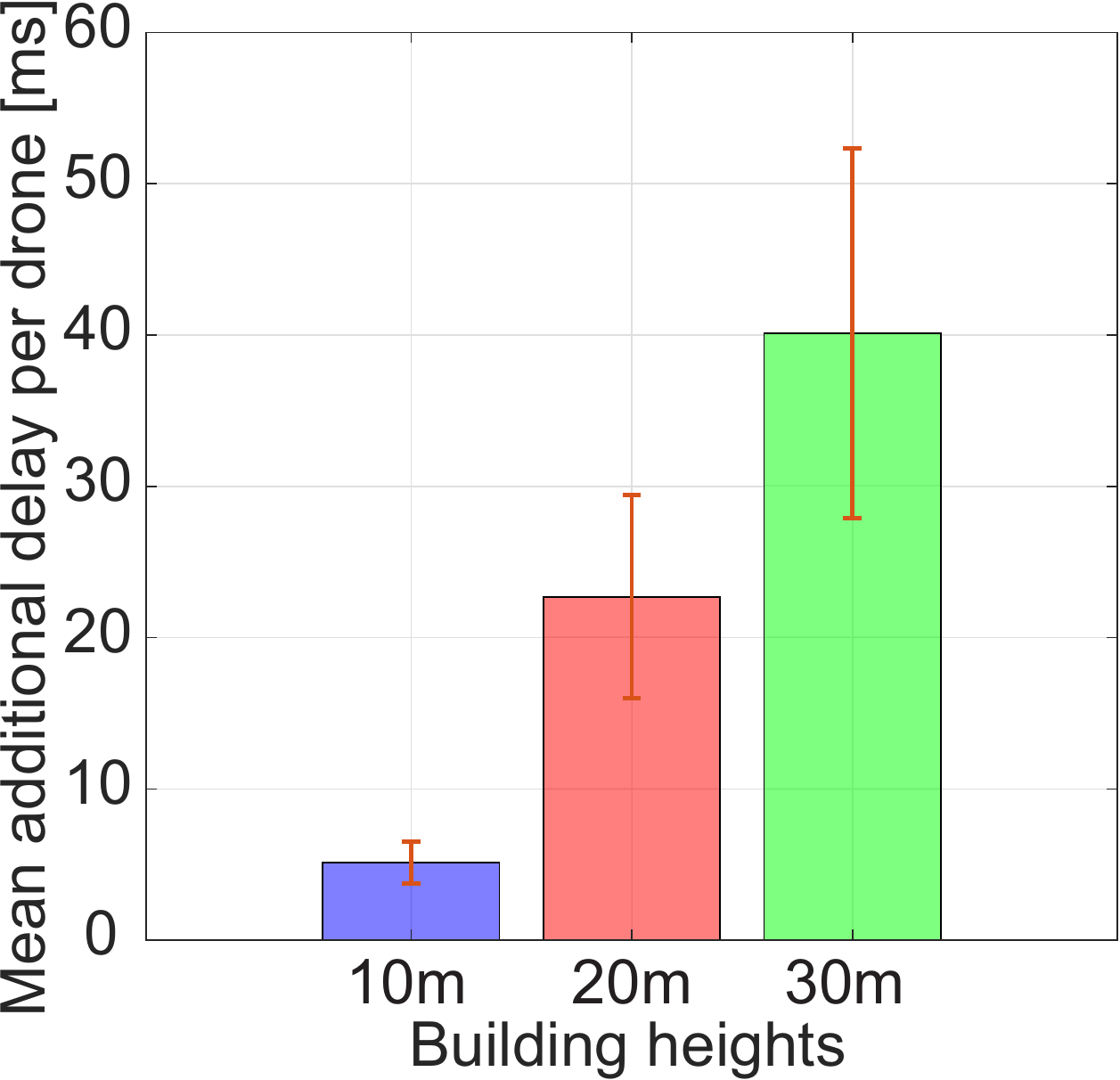}{0.22}{}
	\subfig{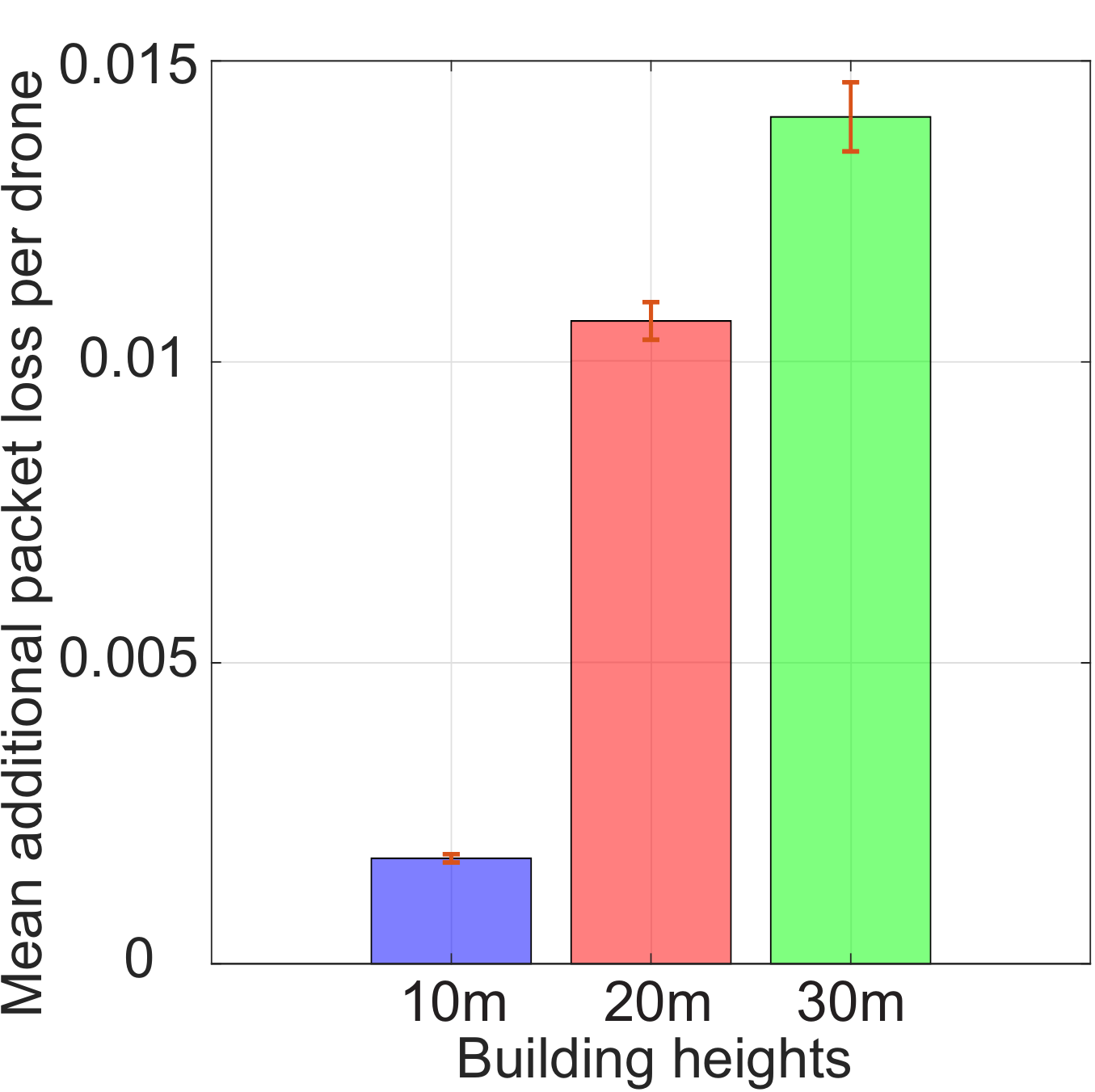}{0.22}{}	
	\caption{Incremental average degradation of the LTE link performance for each added drone with different mean height of buildings.}
	\label{fig:communication-evaluation-height-inc}
\end{figure}

The penalties incurred by the increased drone usage on of the \ac{LTE}-link are detailed further in Fig.~\ref{fig:communication-evaluation-height-inc}. The average impact which adding one more drone to the scenario produces on the delay and \ac{PDR} is analyzed for the different building heights.
On average, going from 1 to 5 drones could incur 150~ms delay, and 50 additional packets out of 1000 may not be delivered correctly within an environment featuring buildings of an average height of 30~m. As opposed to an additional delay of about 20~ms and an extra loss of approximately 5 packets out of 1000 at an average building height of 10~m when increasing the drone count from 1 to 5. 

%\dummyhere{\ac{LTE} at varying building heights}{\ac{LTE} communications at varying building heights}{fig:communication-evaluation-height}

\section{Conclusion}

%
% Introduction
%
In this paper, we investigated the use of relying on \uavs to enhance logistics applications.
%
% Problem statement
%
We analyzed the feasibility of such an enhancement from the network perspective by comparing the performance of \ac{LTE} and \ac{C-V2X} for reliable communications among the delivery fleet.
%
% Solution appraoch
%
The evaluations were carried out using the system-level development platform for joint modeling of aerial and ground-based vehicular networks provided by the open simulation framework \ac{LIMoSim}.
%
% Results
%
It was shown that drone usage in parcel delivery can reduce the time taken to complete a delivery tour by up to 80\%.
Increased drone usage further reduces the completion time and the energy consumption per drone, however, speeding up the tour completion by increasing drone usage has limits.
%
% Future work
%
In a future work, the delivery performance could be enhanced further by using more efficient schemes to schedule delivery jobs for the drones. Such schemes could also consider energy efficiency as a scheduling criterion.
An extension of the presented work could also account for more real-world interactions, such as traffic lights and congestion.
Communication link deterioration around tall buildings still raises a challenge which could, however,  be addressed by leveraging mobility prediction schemes, to forecast the channel conditions at future locations during the path planning procedure of the drone.

\section*{Acknowledgment}

\footnotesize
Part of the work on this paper has been supported by
the German Federal Ministry of Education and Research
(BMBF) in the projects LARUS (13N14133) and
A-DRZ (13N14857) as well as the
Deutsche Forschungsgemeinschaft (DFG) within the Collaborative Research Center SFB 876 ``Providing Information by Resource-Constrained Analysis'', project B4.

\bibliographystyle{IEEEtran}
\bibliography{Bibliography}

\end{document}